\documentclass[twoside]{article}
\usepackage{fleqn,espcrc2}

\usepackage{graphicx}
\usepackage[figuresright]{rotating}

\newcommand{\AmS}{{\protect\the\textfont2
  A\kern-.1667em\lower.5ex\hbox{M}\kern-.125emS}}

\title{Monte Carlo Hamiltonian}

\author{H. Jirari\address{D\'epartement de Physique,
        Universit\'e  Laval, Qu\'ebec, 
        Qu\'ebec G1K 7P4, Canada},
        H. Kr\"oger$^{\rm a}$,
        Chun-Qing Huang\address{Department of 
        Physics, Zhongshan University, Guangzhou 
        510275, China}, 
        Jun-Qin Jiang$^{\rm b}$, 
        X.Q. Luo$^{b}$, and
        K.J.M. Moriarty\address{
        Department of Mathematics, Statistics 
        and Computing Science,     
        Dalhousie University,
        Halifax, Nova Scotia B3H 3J5, 
        Canada}}
       
\begin{document}

\begin{abstract}
We suggest how to construct an effective low energy Hamiltonian via Monte Carlo starting from a given action. We test it by computing thermodynamical observables like average energy and specific heat for simple quantum systems.
\vspace{1pc}
\end{abstract}

\maketitle

\section{INTRODUCTION}

Although most progress in lattice gauge theory has been achieved by use of the Lagrangian formulation, it is worthwhile to keep the Hamiltonian formulation in mind. The potential usefulness of the Hamiltonian lies in the following areas: 
- Non-perturbative computation of {\em scattering cross sections} and 
{\em decay amplitudes} in hadronic systems. 
- {\em Low-lying excited states} of the {\em hadronic spectrum} 
and {\em quantum chaos} in such a system. 
- {\em Hadron wave functions} and {\em hadron structure functions} (for small $x_{B}$ and $Q^{2}$). 
- {\em Finite temperature} and {\em finite density} in baryonic matter (quark-gluon plasma phase transition, neutron stars and cosmology).
- {\em Atomic physics}: study of {\em spectra} and of {\em quantum chaos}. 
- {\em Condensed matter}: study of spin systems (computation of dynamical structure factors), and {\em high $T_{c}$ superconductivity} models (search for electron pair attraction at very small energy).  

Here we suggest to construct an effective low energy Hamiltonian, 
denoted by 
$H_{eff}$ in the following, via Monte Carlo. We suggest to apply the Monte Carlo method for the numerical computation of matrix elements of the propagator, and second for a stochastic selection of states from a basis of Hilbert states.

\section{CONSTRUCTION OF $H_{eff}$}

Let us outline the method, by first assuming that we are given a finite set of Hilbert states, which is sufficiently large and suitably chosen such that 
low energy observables can be well reproduced. 
Later we will discuss the stochastic basis selection. 
We start from a complete orthonormal basis of Hilbert states $| e_{i}>, ~ i=1,2,3 \cdots $ and consider the matrix elements for a given fixed $N$. We consider a transition amplitudes in imaginary time
\begin{equation}
M_{ij}(T) = <e_{i} | e^{-H T /\hbar} | e_{j} >, ~~~ i,j \in 1,\cdots,N ,
\end{equation}
where $H$ is the exact Hamiltonian. If $H$ is Hermitian, then $M$ is unitarily equivalent to a diagonal matrix $D$ such that
\begin{equation} 
M(T) = U^{\dagger} ~ D(T) ~ U .
\end{equation}
By algebraic solution for $D$, we can construct explicitly an effective Hamiltonian
\begin{equation}
H_{eff} = \sum_{k =1}^{N} | E^{eff}_{k} > E^{eff}_{k} < E^{eff}_{k} | ,
\end{equation}
where the $| E^{eff}_{k} >$ are the eigenstates of $D$, and $E^{eff}_{k}$   
are the eigenvalues of $log(D)$. One has the identities
\begin{equation}
U^{\dagger}_{ik} = < e_{i} | E^{eff}_{k} >, ~~~ D_{k}(T) = e^{-E^{eff}_{k} T/\hbar } .
\end{equation}

\section{MATRIX ELEMENTS}

In order to be specific, we work in $D=1$ dimension. We choose basis states $|e_{i}>$ in position space by introducing a lattice with nodes $x_{i}$ and define $e_{i}(x)$ (unnormalized) 
by $e_{i}(x)=1$ if $x_{i} \leq x \leq x_{i+1}$, zero else. 
We choose $\Delta x_{i}=x_{i+1}-x_{i} = \mbox{const}$. The matrix elements read
\begin{eqnarray}
M_{ij}(T) 
&=&  
\int_{x_{i}}^{x_{i+1}} d y 
\int_{x_{j}}^{x_{j+1}} d z 
\nonumber \\
&\times&
\left.
\int [dx] \exp[ - S[x]/\hbar ] \right|^{y,T}_{z,0} .
\end{eqnarray}
Here $S$ denotes the Euclidean action for a given path $C$,
\begin{equation}
S[C] = \left. \int_{0}^{T} dt ~ \frac{1}{2} m \dot{x}^{2} 
+ V(x) \right|_{C} .
\end{equation}
We suggest to compute the matrix elements $M_{ij}(T)$ from the action
via Monte Carlo with importance sampling (Metropolis algorithm). In order to do so we write $M_{ij}(T)$ as a ratio of two integrals by splitting the action
\begin{equation}
S = S_{0} + S_{V} \equiv 
\int_{0}^{T} dt ~ \frac{1}{2} m \dot{x}^{2} + 
\int_{0}^{T} dt ~ V(x) ,
\end{equation}
and expressing $M_{ij}$ by
\begin{eqnarray}
&& M_{ij}(T) = M^{(0)}_{ij}(T) ~
\int_{x_{i}}^{x_{i+1}} d y 
\int_{x_{j}}^{x_{j+1}} d z
\nonumber \\
&& \left. \int [dx] ~ \exp[ - S_{V}[x]/\hbar ] ~ \exp[ -S_{0}[x]/\hbar ] \right|^{y,T}_{z,0}
\nonumber \\
&& / 
\int_{x_{i}}^{x_{i+1}} d y 
\int_{x_{j}}^{x_{j+1}} d z
\nonumber \\
&& \left. \int [dx] ~ \exp[ -S_{0}[x]/\hbar ] \right|^{y,T}_{z,0}  .
\end{eqnarray}
Now we consider $O \equiv \exp[ -S_{V}/\hbar]$ as an observable
and the matrix element $M_{ij}$ is the expectation value of that observable. 
The matrix elements $M^{(0)}_{ij}$, corresponding to the free action $S_{0}$ are known analytically (apart from an overall integration), 
\begin{eqnarray}
&& M^{(0)}_{ij}(T) =  
\int_{x_{i}}^{x_{i+1}} d y 
\int_{x_{j}}^{x_{j+1}} d z ~
\nonumber \\
&& \times \sqrt{ \frac{ m }{ 2 \pi \hbar T } } ~
\exp \left[ - \frac{ m}{2 \hbar T } (y - z)^{2} \right] .
\end{eqnarray}

\section{STOCHASTIC BASIS}

The success of lattice Euclidean lattice calculations depends crucially on the 
efficiency of Monte Carlo with importance sampling when computing the path integral. The path integral is represented by a sum over "representative" configurations (corresponding to themodynamical equilibrium configurations from a Boltzmann distribution). Guided by that we suggest here to construct     
a basis of "representative" states such that the energies of the effective Hamiltonian follow (at least approximatively) a Boltzmann distribution.
The idea is to guide the basis selection by the quantum mechanical Euclidean 
transition amplitude. For the Free system it reads
\begin{eqnarray}
G_{Eucl}(x,T:y,o) &=& \sqrt{\frac{m}{2 \pi \hbar T}} 
\nonumber \\
&\times& \exp[ - \frac{m}{2 \hbar T} (x-y)^{2}].
\end{eqnarray}
This function being positive for all $x,y,T$, can serve as a probability density. We choose $y=0$ and write
\begin{eqnarray}
P(x) &=& \frac{1}{Z} G_{Eucl}(x,T;0,0) ,
\nonumber \\
Z &=& \int dx~G_{Eucl}(x,T;0,0) .
\end{eqnarray}
Then we draw nodes $x_{i_{\nu}}$ randomly, and construct basis states 
as box functions like those of the original nodes $x_{i}$. 
The idea is to work with a stochastic basis of dimension $N_{eff}$ being much smaller than $N$, the dimension of the original basis. This should be feasible in particlular in many-body systems.
In the free case $P(x)$ is a Gaussian,
\begin{equation}
P(x) = \frac{1}{\sqrt{2 \pi } \sigma} \exp[ \frac{x^{2}}{2 \sigma^{2}} ], ~~~ \sigma = \sqrt{\frac{\hbar T}{m}} .
\end{equation}
One can see that this implies a Gaussian distribution for the wave numbers $k$ and hence a Boltzmann distribution for the energies $E$. In the presence of a potential, a possible approximation is to write $G_{Eucl}$ as a path integral retaining only the classical path.

\begin{figure}[htb]
\vspace{3cm}
\caption{Average energy of the harmonic oscillator.
Solid line and diamonds, respectively,
represent the exact analytical result, 
and that from the exact matrix elements for $\Delta x=1$ 
and $N=20$. Fig. taken from Ref.\cite{Jirari99}.}
\label{fig:harmfig1}
\end{figure}
\begin{figure}[htb]
\vspace{3cm}
\caption{Specific heat over $k_B$ of the harmonic 
oscillator. Symbols as in Fig.[1]. Fig. taken from Ref.\cite{Jirari99}.}
\label{fig:harmfig2}
\end{figure}

\section{NUMERICAL RESULTS}

In order to test the Monte Carlo Hamiltonian we have considered the free 
system, the harmonic oscillator and other local potentials, where either analytic or precise approximative solutions are known. We have computed the energy spectrum, wave functions and thermodynamic observables like the partition function $Z$, the average energy $U$ and the specific heat $C$. 
Results for wave functions of the harmonic oscillator are shown in Ref.\cite{Jirari99}. Results for other potentials can be found in Refs. \cite{Huang99,Luo99}. The thermodynamic observables are defined by
\begin{eqnarray}
Z(\beta) &=& Tr[ e^{-\beta H } ] ,
\nonumber \\
U(\beta) &=&  - \frac{ \partial \log Z }{ \partial \beta } ,
\nonumber \\
C(\beta) &=& k_{B} \beta^{2} \frac{ \partial^{2} \log Z } 
{ \partial \beta^{2} } ,
\label{eq:DefThermo}
\end{eqnarray}
where $\beta = (k_{B} {\cal T})^{-1}$, ${\cal T}$ is the temperature, and we identify $\beta$ with the imaginary time $T$ by $\beta = T/\hbar$.
The partition function of the effective Hamiltonian is obtained from its spectrum,
\begin{equation}
Z_{eff}(\beta) = Tr[ e^{- \beta H_{eff} } ] = 
\sum_{k=1}^{N} e^{-\beta E^{eff}_{k}} .
\end{equation}
Via Eq.(\ref{eq:DefThermo}) one obtains the corresponding average energy $U_{eff}$
and the specific heat $C_{eff}$. Fig.[1] shows a plot of the average energy, comparing the exact result with that from the effective Hamiltonian. One observes that the agreement is better where ${\cal T} \to 0$, i.e. in the low energy regime.
A similar behavior is found for the specific heat, shown in Fig.[2].  
In the limit $\beta \to \infty$ the average energy tends to 
the ground state energy, $U \to \hbar \omega/2$ (Feynman-Kac formula).
One should keep in mind that $H_{eff}$ has been constructed for a specific value of the time parameter, $T=1$ corresponding to the temperature ${\cal T} =1$ (we use $\hbar = k_{B}=1$). The effective Hamiltonian, constructed in this way 
describes well thermodynamic observables in the range 
$\beta_{c} \leq \beta$ ~ (it works also for $\beta > 10$, not shown in the figure). 
However, it breaks down for $\beta < \beta_{c}$, i.e. ${\cal T} > {\cal T}_{c}$.
This is due to the small dimension $N=20$ of the matrix. Agreement in a larger $\beta$-region, i.e. lowering $\beta_{c}$ can be obtained by increasing $N$.
The statistical errors are of the size of the symbols in the Figs. 
Here we have not presented any results using the stochastic basis. This is discussed in \cite{Huang99}.

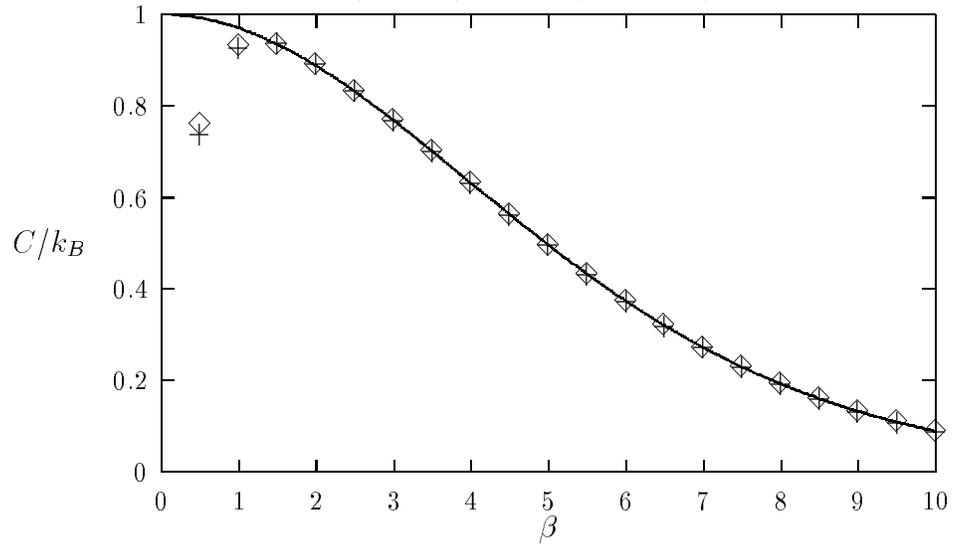
\end{document}